\documentclass[reprint,aps,hypertext]{revtex4-1}

\usepackage[utf8]{inputenc}

\usepackage{amsmath}
\usepackage{amssymb}
\usepackage{hyperref}
\usepackage{color}
\usepackage{booktabs}
\usepackage{tikz}
\usepackage{pgfplots}
\usepackage{verbatim}
\usepackage[english]{babel}

\pgfplotsset{compat=newest}
\usetikzlibrary{external}
\tikzsetexternalprefix{paper-generative-local-figure-}

\usepackage{natbib}

\renewcommand{\deg}{{d}}
\newcommand{\subscripti}{\text{in}}
\newcommand{\subscriptb}{\text{out}}
\newcommand{\wi}{{\lambda_\subscripti}}
\newcommand{\wb}{{\lambda_\subscriptb}}
\newcommand{\alphawi}{{\alpha_\subscripti^+}}
\newcommand{\alphawb}{{\alpha_\subscriptb^+}}
\newcommand{\betawi}{{\alpha_\subscripti^-}}
\newcommand{\betawb}{{\alpha_\subscriptb^-}}
\newcommand{\alphawiprior}{{\alpha^+}}
\newcommand{\alphawbprior}{{\alpha^+}}
\newcommand{\betawiprior}{{\alpha^-}}
\newcommand{\widetildealphawi}{\tilde\alpha_\subscripti^+}
\newcommand{\widetildealphawb}{\tilde\alpha_\subscriptb^+}
\newcommand{\widetildebetawi}{\tilde\alpha_\subscripti^-}
\newcommand{\widetildebetawb}{\tilde\alpha_\subscriptb^-}

\newcommand{\paramwi}{{\alpha_\subscripti}}
\newcommand{\paramwb}{{\alpha_\subscriptb}}
\newcommand{\thetawi}{{\theta_\subscripti}}
\newcommand{\thetawb}{{\theta_\subscriptb}}
\newcommand{\alphad}[1]{{\alpha_{d_#1}}}
\newcommand{\thetad}[1]{{\theta_{d_#1}}}
\newcommand{\thetadall}{{\theta_d}}
\renewcommand{\P}{\mathbb{P}}

\newcommand{\paramwprior}{{\alpha}}
\newcommand{\thetawprior}{{\theta}}
\newcommand{\alphadprior}{{\alpha}}
\newcommand{\thetadprior}{{\theta}}

\newcommand{\within}{w}
\newcommand{\volume}{v}

\newcommand{\GammaPDF}{\Gamma}
\newcommand{\PoissonPDF}{P}

\newcommand{\BetaPDF}{\beta}
\DeclareMathOperator{\expect}{\mathbb{E}}
\DeclareMathOperator{\argmax}{argmax}
\newcommand{\Beta}{B} 

\newcommand{\tblref}[1]{Table~\ref{#1}}
\newcommand{\figref}[1]{Fig.~\ref{#1}}

\newcommand{\secref}[1]{Section~\ref{#1}}

\def\tblskip{\vspace\tblskipamount}
\newskip\tblskipamount
\newcommand{\hlinetop}{\hline\noalign{\tblskip}}
\newcommand{\hlinemid}{\noalign{\tblskip}\hline\noalign{\tblskip}}
\newcommand{\hlinebot}{\noalign{\tblskip}\hline}

\newcommand{\Lapprox}{{\tilde L}}

\newcommand{\YL}{YL}
\newcommand{\HK}{HK}
\newcommand{\PPR}{PPR}

\pgfplotscreateplotcyclelist{datasets}{%
  mark=square*,   blue!90!black, solid, every mark/.append style={solid}\\%
  mark=*,         red!80!black,  solid, every mark/.append style={solid}\\%
  mark=triangle*, green!50!black,  solid, every mark/.append style={solid}\\%
  mark=diamond*,  brown!80!black, solid, every mark/.append style={solid}\\%
  mark=pentagon*, purple!60!black, solid, every mark/.append style={solid}\\%
  mark=star,      teal!60!black, solid, every mark/.append style={solid}\\%
}

\begin{document}

\title{Generative models for local network community detection}
\author{Twan van Laarhoven}
\affiliation{Institute for Computing and Information Sciences, Radboud University Nijmegen, The Netherlands \\
/ Open University, The Netherlands}

%
%

\begin{abstract}

Local network community detection aims to find a single community in a large network, while inspecting only a small part of that network around a given seed node.
This is much cheaper than finding all communities in a network.
Most methods for local community detection are formulated as ad-hoc optimization problems.
In this work, we instead start from a generative model for networks with community structure.
By assuming that the network is uniform, we can approximate the structure of unobserved parts of the network to obtain a method for local community detection.
%
We apply this local approximation technique to two variants of the stochastic block model.
To our knowledge, this results in the first local community detection methods based on probabilistic models.
Interestingly, in the limit, one of the proposed approximations corresponds to conductance, a popular metric in this field.
Experiments on real and synthetic datasets show comparable or improved results compared to state-of-the-art local community detection algorithms.
\end{abstract}

\maketitle 

\section{Introduction}

Networks are a convenient abstraction in many different areas, such as
social sciences,
biology, and the world wide web.
A common structure in these real-world networks is a community, a group of nodes that are tightly connected and often share other properties, for example, biological function in a protein interaction network.

Imagine that you are trying to find such a community of nodes in a network.
If the network is very large, it becomes too expensive to look at all nodes and edges in the network.
Therefore local methods are needed.
%
Local community detection aims to find only one community around a given set of seed nodes, by relying on local computations involving only nodes relatively close to the seed \citep{KloumannKleinberg2014,Andersen2006local}; in contrast to global community detection, where all communities in a network have to be found.
%
%

For global community detection, it is possible to treat the problem of finding all communities in a network as a probabilistic inference problem.
This puts global community detection on a solid foundation, and makes it clear what a community is, and how these communities manifest in the network structure.

But most algorithms for local community detection operate by optimizing an ad-hoc objective function such as conductance \citep{Andersen2006local,Yang2012,Li2015}.

In this paper we will fill this gap, and propose a probabilistic model for local community detection.
%
%
%
%
%
%
%
Our contributions can be summarized as follows:
\begin{enumerate}
  \item We introduce an approximation technique for using global models to perform local community detection.
  \item We introduce the first method for local community detection based on a generative model by using this approximation.
  \item We propose two algorithms for local community detection, based on approximations of the stochastic block model and of the degree-corrected stochastic block model.
  \item We provide a probabilistic interpretation of conductance, as limit behavior of 
     the approximate degree-corrected stochastic block model.
  \item We show that the approximate stochastic block model is a highly competitive algorithm, which outperforms the state-of-the-art on three of five real life benchmark datasets.
\end{enumerate}

\subsection{Related work}

\subsubsection{Local community detection}
Local network community detection methods have largely focused on optimizing conductance, which is a measure of the quality of a graph cut.
Empirically, conductance has been shown to be a good quality metric for communities in real-world networks \citep{Yang2012}, in the sense that real-world communities often have a higher conductance than other sets of nodes. 

Because community detection is computationally hard \citep{Fortunato2010,ShiMalik2000NormalizedCut}, several different heuristics have been developed.
A common approach is to use spectral partitioning, which involves finding the dominant eigenvector of a random walk kernel.
%
\citet{Andersen2006local} showed that communities with a good conductance can be computed efficiently in this way.
In their method nodes are added to the community in order of decreasing personalized pagerank score, and the community along this `{sweep}' with the highest conductance is returned. Computing this personalized pagerank is a global operation, but efficient local approximations are possible, that only involve nodes near the seed.

%
Several variants of this sweep method have been proposed.
\Citet{Kloster2014} propose an alternative to the personalized pagerank score, based on the heat kernel instead of random walks.
\Citet{Yang2012} propose to find the first local optimum of conductance instead of a global optimum.
Other heuristics involve trying to find multiple pagerank-like vectors and restarting the method from different neighborhoods around the seed \citep{Li2015}.

However, being based on graph cuts, a good conductance is often achieved by cutting a network into roughly equal-sized parts, which is undesirable.
To limit the size of communities, a cut-off on the personalized pagerank score can be used \citep{Andersen2006local}; also, variations of the sweep methods have been proposed that stop at earlier local optima \citep{Yang2012}.

%

%



\subsubsection{Global community detection}
Many different global community detection methods have been developed for different classes of networks and different community structures.
For a complete overview, we refer the reader to the surveys by \citet{Fortunato2010,Xie2013OverlappingSurvey}.

%
%

Here we focus on probabilistic models for global community detection.
The simplest are the stochastic block models
\citep{Holland1983sbm,Anderson1992sbm},
which partition the nodes into communities, with varying probabilities of edges.
These block models produce networks that are very different than real networks, in particular, the distribution of degrees is very different.
To more accurately model the node degrees, \citet{Karrer2011} have proposed the degree-corrected stochastic block model (DC-SBM).
In this model, they add an extra parameter to each node, which controls the likelihood of edges to that node, and hence the node's degree.
This extra complexity comes at the cost of making the model more difficult to fit, and so degree correction might not be appropriate for all networks \citep{Yan2014DCBM}.


An issue with the stochastic block model is that the number of communities has to be fixed because with more communities there are more parameters in the model, which makes it impossible to compare likelihoods. The Infinite Relational Model \citep{Kemp2006IRM} solves this problem by assuming an infinite number of communities in combination with a Chinese restaurant prior over community structures.

The model of \citet{NewmanLeicht2007mixture} goes one step further, and has a parameter for each combination of node and community, indicating the likelihood of edges from nodes in that community to a particular other node. This is similar to models based on non-negative matrix factorization \citep{Psorakis2011NMF,Ball2011}. These more complex models allow for nodes to be in more than one community.

Aside from these flat models, also hierarchical models have been developed for probabilistic network community detection
\citep{Zhang2007,Blundell2013BayesianHierarchical}.

%
%
%
%
With all probabilistic models there is the question of inference, that is, how to find the parameters or distribution of parameters that accurately model the observed data.
Common approaches are to use maximum likelihood \citep[used by e.g.][]{Karrer2011,Yan2014DCBM},
variational Bayes \citep[used by][]{Hofman2008}, and Markov chain Monte Carlo sampling \citep[used by e.g.][]{Kemp2006IRM}.

More recently, there has been work using loopy belief propagation for inference in stochastic block models \citep{Decelle2011BeliefPropagationBlockModel,ZhangMooreNewman2016UnequalGroups}.
There it has also been noted that these models exhibit a phase transition: beyond a certain point in parameter space it is impossible to recover the true community structure, even it was drawn from the model itself. Furthermore, this transition is sharp for large networks.


%

\subsection{Problem description}

Before continuing, we will formalize the problem.
The network of interest is represented as an unweighted undirected graph without self-loops.
Let $N$ be the number of nodes in this graph, and $M$ the number of edges.
The graph can be represented by an adjacency matrix $A$, where $a_{ij}=a_{ji}=1$ if there is an edge between nodes $i$ and $j$, and $a_{ij}=0$ otherwise.
Furthermore $\sum_{i=1}^N \sum_{j=1}^N a_{ij} = 2M$.
%

The local community detection problem is now to find the community $c_s$ that contains a given seed node $s$,
while inspecting only nodes and edges in or near that community.
We will only concern ourselves with a single seed node in this paper, but this is not essential.
%
%

When working in a probabilistic setting, the goal becomes to find the most likely community that contains the seed, $\argmax_{c_s} \P(c_s \mid A,s)$.
Even computing this probability entails a marginalization over all the other clusters in the graph.
So as a first simplification we will instead search for the most likely clustering, $\argmax_{C} \P(C \mid A,s)$, and report the community $c_s$ in this clustering.

We can assume that the seed is chosen independently from the graph, and also independently from the clustering, so we have that
$
  \P(C \mid A,s) \propto \P(C) \P(A \mid C).
$ 
To find the community containing the seed we only need to maximize this quantity, and then to find the community $c_s \in C$ that contains the seed.


\section{The Stochastic Block Model}


We first model the community structure of all nodes in the graph using
the stochastic block model \citep{Karrer2011}.
%
%
In this global model, each node $i$ is in exactly one community, which we denote as $c_i$, so the communities form a partition of the set of nodes.
The edges of the graph are generated independently based on these communities.
In the standard stochastic block model the probability of an edge between nodes $i$ and $j$ is $\P(a_{ij}) = \pi_{c_ic_j}$ where the $\pi$ are parameters.

For simplicity, we only consider two different values for $\pi$, $\pi_{c,c}=\wi$ for edges inside clusters and $\pi_{cc'}=\wb$ for edges between different clusters $c \neq c'$, so
\begin{align*}
   \P(a_{ij}) =\pi_{c_ic_j}
   &= \begin{cases}
             \wi & \text{ if } c_i = c_j \\
             \wb & \text{ if } c_i \neq c_j.
           \end{cases}
\end{align*}
Since we do not know the value of these parameters $\wi$ and $\wb$, we put a conjugate Beta prior on them,
\begin{align*}
  \wi,\wb &\sim \BetaPDF(\alphawiprior,\betawiprior).
\end{align*}

In this simpler variant of the stochastic block model, the number of parameters of the model does not depend on the number of communities.
Hence, in contrast to most other work on stochastic block models, we do not have to fix the number of communities.
Instead, we use a prior over partitions, which allows for varying number of communities of varying sizes.

It is well known that community sizes in real life networks follow a power law distribution \citep{Palla2005-clique-percolation}. Hence we adopt the prior
\begin{align}
  \label{eq:powerlaw}
  \P(C) \propto \sum_{c \in C} (\gamma-1)|c|^{-\gamma},
\end{align}
where $C$ ranges over all possible partitions of the set of $N$ nodes.
Note that the particular choice of prior distribution is not critical to the rest of this work, and for other applications other priors might make sense. In particular, a common alternative choice is the Chinese Restaurant Process.


\subsection{Inference}

In the basic stochastic block model
\begin{multline*}
  \P(A \mid C)
    = \expect_{\wi,\wb}\Bigl[ \prod_{i<j} \pi_{c_ic_j}^{a_{ij}} (1-\pi_{c_ic_j})^{1-a_{ij}} \Bigr]
   \\ = \expect_{\wi,\wb}\Bigl[
       \wi^{\alphawi} (1-\wi)^{\betawi}
       \wb^{\alphawb} (1-\wb)^{\betawb}
       \Bigr],
\end{multline*}
where $\alphawi = \sum_{i<j,c_i=c_j} a_{ij}$, $\betawi = \sum_{i<j,c_i=c_j} (1-a_{ij})$ and $\alphawb$ and $\betawb$ are the corresponding sums over $c_i \neq c_j$.
This likelihood has the same shape as the beta distribution, so we can calculate the expectation exactly, to get
\begin{multline}
  \label{eq:sbm-lh}
  \P(A \mid C) = \\ \quad
  \frac{\Beta(\alphawiprior+\alphawi,\betawiprior+\betawi)\Beta(\alphawbprior+\alphawb,\betawiprior+\betawb)}{\Beta(\alphawiprior,\betawiprior)^2}.
\end{multline}
Multiplying this by the prior on clusterings \eqref{eq:powerlaw} gives the posterior probability $\P(C|A)$ up to a normalizing constant.

\subsection{Local approximation}
\label{sec:asbm}

The likelihood in equation~\eqref{eq:sbm-lh} is still a function of the entire clustering, so to find the most likely cluster containing the seed we would need to consider clusterings of the entire graph.
To obtain a local model we make an approximation based on the assumption that the graph is uniform: all clusters of the graph are similar to each other.

We make this idea concrete by assuming that all clusters in the graph have approximately the same volume, the same size, and the same fraction of within community edges.
%
Now, if the community containing the seed has $n$ nodes, while the graph has $N$ nodes, this means that there are approximately $k=N/n$ communities that are all similar to $c_s$.
Furthermore, suppose that this community has $w$ within community edges, then 
the parameters for the stochastic block model can be approximated as
\begin{align*}
  \widetildealphawi &= k w, \\
  \widetildebetawi  &= k n(n-1)/2 - \widetildealphawi,\\ 
  \widetildealphawb &= M - \widetildealphawi, \\
  \widetildebetawb  &= N(N-1)/2 - \widetildealphawi - \widetildebetawi -\widetildealphawb.
\end{align*}
With these quantities we can use equation~\eqref{eq:sbm-lh} to approximate the likelihood of the network given the community that contains the seed. And hence also approximate the posterior probability of a community given the network,
\begin{multline}
  \tilde P_\text{SBM}(c_s,C \mid A) =
  (\gamma-1)^kn^{-k\gamma}
  \\ \frac{\Beta(\alphawiprior+\widetildealphawi,\betawiprior+\widetildebetawi)\Beta(\alphawbprior+\widetildealphawb,\betawiprior+\widetildebetawb)}{\Beta(\alphawiprior,\betawiprior)^2}
  .
\end{multline}

Note that instead of taking $k=N/n$, we might reason that, if the volume of the community containing the seed is $v$ and the graph has $M$ edges, that there are approximately $k=2M/v$ communities. This is, in general, a different estimate. For the stochastic block model, it makes more sense to use $k=N/n$, because then there is no dependence on the volume of the community.


\section{Degree-corrected block model}

The degree distribution of the stochastic block model is not very realistic, because nodes inside a cluster have similar degrees.
In many real-world networks, there are hub nodes, which have a much higher than average degree; as well as leaf nodes with a very low degree.
To accurately model these phenomena \citet{Karrer2011} have proposed the degree-corrected stochastic block model (DC-SBM).
In this model, they assign an extra parameter $\deg_i$ to each node, which controls the likelihood of edges to that node, and hence the node's degree.
%
%

We can then model the edges as being drawn from a Poisson distribution with mean $\deg_i \deg_j \pi_{c_ic_j}$,
\begin{align*}
  a_{ij} &\sim \PoissonPDF(\deg_i \deg_j \pi_{c_ic_j})  \text{ for all } i < j.
\end{align*}
Note that we use a Poisson distribution, which allows for weighted edges with weight larger than 1, instead of the Bernoulli distribution, because the mean might be larger than $1$.



We again place conjugate priors on all parameters, which in this case follow a gamma distribution,
\begin{align*}
  \wi,\wb &\sim \GammaPDF(\paramwprior,\thetawprior) \\
  \deg_i  &\sim \GammaPDF(\alphadprior,\thetadprior) \text{ for all } i.
\end{align*}
%


\subsection{Inference}
%
%
%


For this degree-corrected model the likelihood of the network given the clustering depends on parameters $\deg$ and $\lambda$.
It is not possible to integrate over these parameters analytically,
and other authors have therefore chosen to maximize over the parameters instead \citep{Karrer2011,Yan2014DCBM}.
Here we use a variational approximation,
%
\begin{multline}
  \log \P(A \mid C)
   \ge 
     L(A,C) \\=
     \expect_{\deg,\lambda \sim Q}\bigl[\log \P(A,\deg,\lambda \mid C) - \log Q(\deg,\lambda)\bigr]
   .
\end{multline}
As is standard, we take $Q$ to be the factorized distribution
$Q(\deg,\lambda) = Q_{\wi}(\wi)Q_{\wb}(\wb) \prod_{i=1}^N Q_{\deg_i}(\deg_i),$
where each component has a gamma distribution,
$Q_{\deg_i}(\deg_i) = \GammaPDF(d_i;\alphad{i},\thetad{i})$, 
$Q_{\wi}(\wi) = \GammaPDF(\wi;\paramwi,\thetawi)$.
This gives us the following variational lower bound
\begin{multline}
  L(A,C)
    = 
     \sum_{i<j} \bigl(a_{ij}\log(\deg_i \deg_j \pi_{c_ic_j}) -\deg_i \deg_j \pi_{c_ic_j} \bigr)\\
    - \sum_{i=1}^N D_{KL}(\alphad{i},\thetad{i} || \alphadprior,\thetadprior)
     \\
    - D_{KL}(\paramwi,\thetawi  || \paramwprior,\thetawprior) \\
     - D_{KL}(\paramwb,\thetawb || \paramwprior,\thetawprior)
    .
  \label{eq:vb}
\end{multline}
Where
$D_{KL}(\alpha,\theta || \alpha',\theta')$
is the Kullback-Leibler divergence between two gamma distributions.
And we have assumed that $a_{ij} \in \{0,1\}$, which implies that $a_{ij}!=1$.

We find that the parameters that maximize $L(A,C)$ are
\begin{align*}
  \alphad{i} &= \alphadprior - 1 + \sum_{j=1}^N a_{ij} \\
  \thetad{i} &= \Bigl( \thetadprior^{-1}  + \sum_{j\neq i} \pi_{c_i,c_j}\alphad{j}\thetad{j} \Bigr)^{-1}
\end{align*}
and
\begin{align*}
  \paramwi &= \paramwprior - 1 + \sum_{i<j,c_i=c_j} a_{ij}
  \\
  \thetawi &= \Bigl( \thetawprior^{-1} + \sum_{i<j,c_i=c_j} \alphad{i}\thetad{i}\alphad{j}\thetad{j} \Bigr)^{-1},
\end{align*}
and similarly for $\paramwb$ and $\thetawb$.
There is a mutual dependence between these variables, so in practice we use several iterations of the above equations to find a good approximation of the parameters.

%


%
%
%

The variational approximation gives a lower bound to the log-likelihood $\P(A \mid C)$.
In contrast, maximum likelihood would give an upper bound. This upper bound is similar to $L(A,C)$, but it does not include the Kullback-Leibler terms. 
For large networks the first term of $L(A,C)$ dominates, and so the variational lower bound, the true likelihood and the maximum likelihood upper bound will all be close.

\subsection{Local approximation}

As before, we will make a local approximation of $L$, which depends only on the community that contains the seed.

\newcommand{\hatv}{\hat \volume}
\newcommand{\hatu}{\hat K}
\newcommand{\hatM}{\hat M}
It will be convenient to define
\begin{align*}
\hatv &= \sum_{i \in c_s} \alphad{i} = \within+n(\alphadprior-1),\\
\hatM &= \sum_{i=1}^N \alphad{i} = 2M+N(\alphadprior-1), \text{ and}\\
\hatu^2 &= \sum_{i \in c_s} \alphad{i}^2.
\end{align*}

First of all, we can approximate $\thetad{i}$ by changing the sum over all other nodes $j\neq i$ to a sum over all nodes $j$.
Then $\thetad{i}$ becomes the same for all nodes in the same community, and under the assumption that all communities are the same, $\thetad{i}=\thetadall$ is the same for all nodes,
%
%
%
%
\begin{align*}
  \thetadall 
             = \Bigl( \thetadprior^{-1} + \thetadall\wi\hatv +\thetadall\wb(\hatM-\hatv) \Bigr)^{-1}.
\end{align*}


Furthermore, because $\alphad{i}$ does not depend on the clustering, we can make the following approximation $\Lapprox$ of $L$,
\begin{multline}
  \label{eq:lapprox}
  L(A,C) \approx \Lapprox(A,c) = 
  2M \log(\thetadall)
  + k\within (\psi(\paramwi)+\log(\thetawi))
  \\
  + (M-k\within)(\psi(\paramwb)+\log(\thetawb))
  \\
  - k(\hatv^2 - \hatu^2) \thetadall^2 \paramwi\thetawi
  - (\hatM^2 - k\hatv^2) \thetadall^2 \paramwb\thetawb
  \\
  + N \alphadprior\log\thetadall - \hatM\thetadall/\thetadprior
  \\
  - D_{KL}(\paramwi,\thetawi || \paramwprior,\thetawprior)
  - D_{KL}(\paramwb,\thetawb || \paramwprior,\thetawprior)
  + \kappa,
\end{multline}
where $\kappa$ is a constant that depends only on the network and on the priors.

The likelihood of the degree-corrected model is based on the degrees of nodes and the volume of communities.
So in contrast to the previous section, here it makes sense to estimate the number of communities as $k=2M/v$ instead of $N/n$.

As before, we multiply this approximate likelihood by the prior, which we approximate as $\P(C)\approx (\gamma-1)^k n^{-k\gamma}$, to obtain the posterior
\begin{align*}
 \tilde P_\text{DCBM}(c_s, C \mid A) = e^{\Lapprox(A,c)} (\gamma-1)^k n^{-k\gamma}.
\end{align*}

\section{Limiting behavior}
\label{sec:limiting-behavior}

The premise of local community detection is to find a community without considering the entire graph, or even a significant portion of the graph.
This is only possible if the community is small compared to the graph.
We can take this assumption one step further, and consider what happens if the graph becomes infinitely large compared to the cluster.

Therefore we take the limit of the approximate likelihood as $N \to \infty$, assuming that the average degree $M/N$ remains constant.
With the stochastic block model from \secref{sec:asbm} we get that
\begin{equation*}
  \lim_{N \to \infty} \frac{\log \tilde P_\text{SBM}(c_s,C\mid A)}{N \log N} = \frac{\within}{n}.
\end{equation*}

With the degree-corrected model we obtain
\begin{equation*}
  \lim_{N \to \infty} \frac{2 \log \tilde P_\text{DCBM}(c_s,C\mid A)}{N \log N} =  \frac{\within}{\volume} - 1,
\end{equation*}
which is exactly equal to the negation of conductance.
In other words, under this model, and in the limit of an infinitely large graph, the a posteriori most likely cluster corresponds to the cluster of minimum conductance.


Note that conductance has a global optimum with a large community that contains all nodes, since in that case $w=v$.
Even in the non-limiting case, as the network becomes larger, so does the optimal community. And for very large networks it becomes impossible to recover small communities. This phenomenon is called the resolution limit \citep{Fortunato2007ResolutionLimit}, and is shared by many network community detection methods.
To avoid the resolution limit, a parameter must be introduced into the objective function, for instance by replacing the graph size or graph volume \citep{Reichardt2004}. In our case, we could take $N$ as a formal parameter, instead of using the actual number of nodes in the network (keeping the average degree $M/N$ fixed). In this way, the search for a community is in effect performed in a subnetwork of a given size.

\section{Experiments}

In this section, we experimentally evaluate the proposed models and approximations. We use the following experimental protocol:
%

\begin{enumerate}
  \item pick a random community from the set of all communities.
  \item pick a random seed from this community.
  \item run the method(s) with this seed.
  \item compare the recovered community to the true one using the $F_1$ score.
  For sets of nodes $c$, $d$ the $F_1$ score amounts to 
  $
    F_1(c,d) = 2|c\cap d|/(|c|+|d|),
  $
  which is $1$ if the communities are identical, and $0$ if they are disjoint.
  We exclude the seed from this comparison since it always occurs in both communities, and we would otherwise see a good $F_1$ score for the trivial community containing only the seed.
\end{enumerate}

\subsection{Methods}

We compare three classes of methods.

\subsubsection{Global generative models}
We optimize the likelihood of the global clustering models with a Louvain method \citep{Blondel2008}.
This yields a partition of the nodes.
In this partition, there is always a single community that contains the seed.
We denote this method as gSBM (global stochastic block model) and gDCBM (global degree-corrected block model).
We use uninformative priors for all parameters, $\BetaPDF(1,1)$ in the stochastic block model and $\Gamma(1,1)$ in the degree-corrected model.
For the power law prior on community sizes we use $\gamma=2$. 

Note that it is somewhat unfair to compare local and global models.
A global method has access to more information. The goal of local community detection is not to outperform global methods, but rather to achieve comparable results faster while looking at only a small part of the network.

\subsubsection{Local approximations}
We have implemented a simple greedy algorithm to optimize $\tilde \P(C\mid A)$ for the stochastic and degree-corrected block models.
The algorithm starts from the community $\{s\}$ that contains only the seed.
Then we consider all neighboring nodes of the current community in a random order, and for each node we add it to the community if doing so would improve the approximate likelihood.
This optimization procedure is then repeated, until none of the neighboring nodes are added.
We further restart this search 10 times, and pick the community with the highest approximate likelihood.
We denote this method as aSBM (local approximate stochastic block model) and aDCBM (local approximate degree-corrected block model).

Each iteration of this greedy optimization procedure takes time proportional to the volume of the retrieved community, and the total number of iterations is bounded by the diameter of the community $D$, which is very small in practice. This makes the total runtime $O(vD)$.

We also consider a variant of aDCBM with an explicit parameter $N$, as discussed in \secref{sec:limiting-behavior}.
We report results with $N=1000$, and set $M$ to the average node degree times $N$ (aDCBM1k).
The supplementary material includes results for different values of $N$.

\subsubsection{State-of-the-art methods for local community detection}
\begin{itemize}
  \item \PPR. The algorithm by \citet{Andersen2006local} based on the Personalized Page Rank graph diffusion.
  We use the implementation included with the {\HK} method.

  \item \HK.  The algorithm by \citet{Kloster2014}, using a Heat Kernel.
   Code is available at \url{https://www.cs.purdue.edu/homes/dgleich/codes/hkgrow}.
  
  \newcommand{\score}{\phi}
  \item \YL.  The algorithm by \citet{Yang2012} with conductance as scoring function.
  This method uses a different stopping condition compared to \PPR, selecting a local optimum of conductance, instead of searching for a more global optimum. This introduces a bias towards finding smaller communities.
  %
  %
  
  \item LEMON.  The Local Expansion via Minimum One Norm algorithm by \citet{Li2015}.
  Instead of considering a single probability vector as in \HK\, or \YL\,, this method uses the space spanned by several short random walks. Communities are found by solving an $l_1$-penalized linear programming problem.
  The algorithm includes a number of  heuristic post-processing steps.
  Code is available at \url{https://github.com/yixuanli/lemon}.
\end{itemize}

\subsection{Artificial datasets}

We first look at artificial datasets,
by using the LFR benchmark \citep{Fortunato2008BenchmarkGraphs} to generate networks with a known community structure.
We used the parameter settings \texttt{N=5000 k=10 maxk=50 t1=2 t2=1 minc=20 maxc=100}, which means that the graph has 5000 nodes,
and between 20 and 100 communities, each with between 10 and 50 nodes.
%
We vary the mixing parameter (\texttt{mu}), which determines what fraction of the edges are between different communities. More mixing makes the problem harder.

The LFR benchmark is very similar to the degree-corrected block model. The differences are that node degrees follow a power law distribution in the LFR model, while we used a gamma distribution, and that in the LFR benchmark the edges are not completely independent because the degree of each node must match a previously drawn value. Nevertheless, we expect the DCBM to give a good fit to these networks.


The results of these experiments are shown in \tblref{tbl:f1}.
Here we see that the global degree-corrected model performs better than the simple stochastic block model.
This is not surprising, since the LFR benchmark has nodes with varying degrees.
These results carry over to the local approximations, which do perform significantly worse than the global models on these datasets.
Out of the local methods, the aDCBM and LEMON models achieve the best results.


%



\subsection{Real-world networks}

\begin{table}
  \caption{
    Overview of the SNAP datasets used in the experiments.
  }
  \label{tbl:datasets}
  \small
  \begin{center}
\begin{tabular}{lrr@{\hspace*{5mm}}r@{\hspace*{3mm}}r}
\hlinetop
Dataset & \#node & \#edge & \#comm & $\overline{|c|}$ \\
\hlinemid
Amazon      &  334863 &   925872 & 151037 & 19.4\\
DBLP        &  317080 &  1049866 &  13477 & 53.4\\
Youtube     & 1134890 &  2987624 &   8385 & 13.5\\
LiveJournal & 3997962 & 34681189 & 287512 & 22.3\\
Orkut       & 3072441 & 117185083 & 6288363 & 14.2\\
\hlinebot
\end{tabular}
  \end{center}
\end{table}

We use five social and information network datasets with ground-truth from the SNAP collection \citep{snapnets}.
These datasets are summarized in \tblref{tbl:datasets}.
We consider all available ground-truth communities with at least 3 nodes.
All experiments were performed on a random subsample of 1000 communities.

\Citet{Yang2012} also defined a set of top 5000 communities for each dataset. These are communities with a high combined score for several community goodness metrics, among which is conductance.
We therefore believe that communities in this set are biased to be more easy to recover by optimizing conductance.

In addition to the SNAP datasets, we also include the Flickr social network \citep{Wang-etal12-flickr-dataset}.
As well as some classical datasets with known communities:
Zachary's karate club \cite{Zachary1977};
Football: A network of American college football games \citep{GirvanNewman2002};
Political books: A network of books about US politics \citep{Krebs2004polbooks}; and
Political blogs: Hyperlinks between weblogs on US politics \citep{Adamic2005polblogs}.
These datasets might not be very well suited for local community detection since they have very few communities.
%

%

%

We see in \tblref{tbl:f1} that, while on the artificial benchmark networks the global model significantly outperforms the local approximation,
on the real-world networks this is not the case. 
We believe that this is because the ground-truth communities on these networks are much smaller, and the considered local methods tend to find smaller communities.

Additionally, the simple stochastic block model outperforms the degree-corrected model on all SNAP datasets except for the Amazon dataset. We found this surprising, because all these datasets do have nodes with widely varying degrees.
However, the number of within community edges varies much less. For instance a node in the DBLP dataset with degree $d_i$ will have on the order of $\sqrt{d_i}$ within community edges, which means that the truth is in between the aSBM (which assumes $O(1)$ edges) and aDCBM models (which assumes $O(d_i)$ edges).

An issue with the local approximation is that
nodes inside communities are not representative of the entire network.
For instance, on the Youtube network the average node degree is $5.3$, while the average degree of nodes that are inside at least one community is $33.7$.

When using an explicit $N=1000$, the results on the SNAP networks improve, again because the ground-truth communities on these networks tend to be small.
For the three largest datasets, different values of $N$ have a large influence on the size of the recovered community. This is likely due to the fact that it is possible to find communities at all scales in these networks.
See the supplementary material for the results for different values of $N$.

The results using the top 5000 communities are much better, which is not surprising, since these top communities were selected to be easier to find.
The trend between the different methods is similar, with the aSBM and aDCBM methods performing best in most cases.

Surprisingly, the local approximations outperform the global community detection methods on most of the real-world datasets.
The likely reason is that the ground-truth communities in the SNAP datasets are relatively small.
And because of the greedy optimization strategy, the local methods tend to find smaller communities.
The global methods, in contrast, find larger communities that better fit the model, but which likely combine several ground-truth communities.
This means that the local methods achieve a much better precision at the cost of a somewhat lower recall, resulting in an overall higher $F_1$ score.

Results on the smaller networks are mixed. All these datasets, except for football have very large clusters for their size, which are easier to recover with the HK and PPR methods. The aDCBM method with $N=1000$ is also sometimes better able to find good communities on these datasets, because those networks have fewer than 1000 nodes, so increasing $N$ increases the size of the found community.

We were unable to run LEMON on the large SNAP datasets due to its memory usage.

\begin{table*}[t]
  \caption{
    $F_1$ score between recovered communities and ground-truth (excluding the seed node).
    The best result for each dataset is indicated in bold, as are the results not significantly worse according to a paired T-test (at significance level $0.01$).
  }
  \label{tbl:f1}
  \small
  \begin{center}
\begin{tabular}{l|cc|ccc|cccc}
\hlinetop
 & \multicolumn{2}{|c|}{Global methods} & \multicolumn{7}{|c}{Local methods} \\
\noalign{\tblskip}
Dataset & gSBM & gDCBM & aSBM & aDCBM & aDCBM-1k & {\YL} & LEMON & {\HK} & {\PPR} \\
\hlinemid
LFR (mu=0.1) & 0.999 & \textbf{1.000} & 0.613 & 0.911 & 0.895 & 0.307 & 0.925 & 0.881 & 0.352\\
LFR (mu=0.2) & 0.998 & \textbf{1.000} & 0.583 & 0.812 & 0.799 & 0.274 & 0.859 & 0.090 & 0.136\\
LFR (mu=0.3) & 0.958 & \textbf{0.997} & 0.534 & 0.800 & 0.726 & 0.168 & 0.587 & 0.039 & 0.040\\
LFR (mu=0.4) & 0.920 & \textbf{0.990} & 0.466 & 0.659 & 0.529 & 0.121 & 0.533 & 0.039 & 0.040\\
LFR (mu=0.5) & 0.756 & \textbf{0.911} & 0.368 & 0.458 & 0.322 & 0.085 & 0.427 & 0.039 & 0.041\\
LFR (mu=0.6) & \textbf{0.433} & \textbf{0.426} & 0.258 & 0.138 & 0.093 & 0.069 & 0.279 & 0.037 & 0.039\\
\hlinemid
Amazon & 0.330 & 0.245 & 0.395 & \textbf{0.431} & \textbf{0.447} & 0.381 & 0.229 & 0.221 & 0.119\\
DBLP & 0.287 & 0.220 & \textbf{0.406} & 0.349 & 0.344 & 0.245 & 0.240 & 0.199 & 0.194\\
Youtube & 0.040 & 0.054 & \textbf{0.099} & 0.071 & 0.084 & 0.082 & 0.075 & 0.031 & 0.052\\
LiveJournal & 0.041 & 0.025 & \textbf{0.072} & 0.043 & 0.052 & 0.054 & - & 0.028 & 0.035\\
Orkut & 0.010 & 0.007 & 0.020 & 0.014 & 0.016 & \textbf{0.034} & - & \textbf{0.032} & 0.019\\
\hlinemid
Amazon (top 5000) & 0.792 & 0.614 & 0.844 & \textbf{0.903} & \textbf{0.895} & 0.780 & 0.397 & 0.709 & 0.527\\
DBLP (top 5000) & 0.442 & 0.334 & \textbf{0.647} & 0.567 & 0.571 & 0.419 & 0.339 & 0.342 & 0.329\\
Youtube (top 5000) & 0.083 & 0.085 & 0.140 & 0.195 & 0.172 & \textbf{0.241} & 0.098 & 0.067 & 0.116\\
LiveJournal (top 5000) & 0.507 & 0.354 & 0.666 & \textbf{0.715} & 0.672 & 0.521 & - & 0.569 & 0.478\\
Orkut (top 5000) & 0.233 & 0.195 & \textbf{0.334} & \textbf{0.312} & 0.119 & 0.097 & - & \textbf{0.312} & 0.260\\
\hlinemid
Karate & 0.165 & 0.101 & 0.379 & 0.448 & 0.740 & 0.562 & 0.683 & 0.799 & \textbf{0.908}\\
Football & \textbf{0.790} & \textbf{0.817} & 0.727 & 0.682 & 0.769 & \textbf{0.784} & 0.322 & 0.452 & 0.260\\
Pol.Blogs & 0.192 & 0.039 & 0.103 & 0.040 & 0.090 & 0.015 & 0.151 & \textbf{0.661} & 0.535\\
Pol.Books & 0.274 & 0.429 & 0.295 & 0.243 & 0.451 & 0.175 & 0.605 & \textbf{0.629} & \textbf{0.653}\\
Flickr & \textbf{0.204} & 0.090 & 0.164 & 0.050 & 0.066 & 0.012 & 0.025 & 0.054 & 0.118\\
\hlinebot
\end{tabular}
  \end{center}
\end{table*}


%


\section{Discussion}





The local approximations are based on the assumption that all communities are alike. We needed to make this assumption to be able to say something about the global properties of the network given only a single community.
In real-world networks there is often a large variation in the size of the communities. Our approximation might work well if the community is close to the average size, but for very large or very small communities it is not accurate.
Better approximations might be possible by finding more than one community (but still a small subset of the network), or by modeling the distribution of community sizes.

When we are interested in local community structure, it would seem to make sense to consider models that only have this local structure. For instance, models with a single community that stands apart from a background.
But to fit such a model to an observed graph we would also need a good model for the background, and so we should also model the structure in the background. In other words, we also need to find communities in the rest of the graph, and the method would not be local.
If we were to instead use a background without further structure, then the closest fit to the observed network will be obtained by using the structure of the single community to explain the largest variances in the entire network, so the obtained `community' would cover roughly half of the nodes. Our approach of assuming that the background is similar to the community containing the seed is a good compromise, as illustrated by the experiments.

In this work we maximize over clusters. It would be interesting and useful to estimate the marginals instead.
That is, the probability that a node $i$ is inside the same cluster as the seed, conditioned on the graph.
While Variational Bayes gives a decent approximation to the log-likelihood, it does not give approximations to the marginals. 
Indeed, trying to use a variational bound to marginalize over all but one of the cluster membership indicators leads to an objective that is identical to $\Lapprox$, except for a relatively small entropy term.
It remains to be seen if other inference methods can be used to estimate the marginals, and thus to in some sense find \emph{all} possible communities containing a given seed.

The assumption used to derive the local approximations is not particular to the stochastic block models that we have used here, and the same technique can be used for other global community detection methods that are based on a partition of the nodes, such as the models of \citet{Kemp2006IRM} or \citet{NewmanLeicht2007mixture}.
However, it is often assumed that in practice some nodes can belong to more than one community.
There exist several global models that include overlapping communities \citep[see e.g.][]{McDaid2010moses,Ball2011}.
In the derivation we used the fact that if all communities are identical and each node is in exactly one community, then there are $N/n$ communities. But when nodes can be in more than one community, it is no longer clear how many communities there are.
We would need a reliable estimate of the average number of communities that cover a node.
It therefore remains an open problem how the local approximation can be applied to models with overlapping clusters.



\section*{Acknowledgements}

This work has been partially funded by the Netherlands Organization for Scientific Research (NWO) within the EW TOP Compartiment 1 project 612.001.352.


\bibliography{clustering}
\bibliographystyle{apsrev4-1}

\appendix
\section{Additional results}
%

\begin{figure*}[ht]
  \centering
  \includegraphics{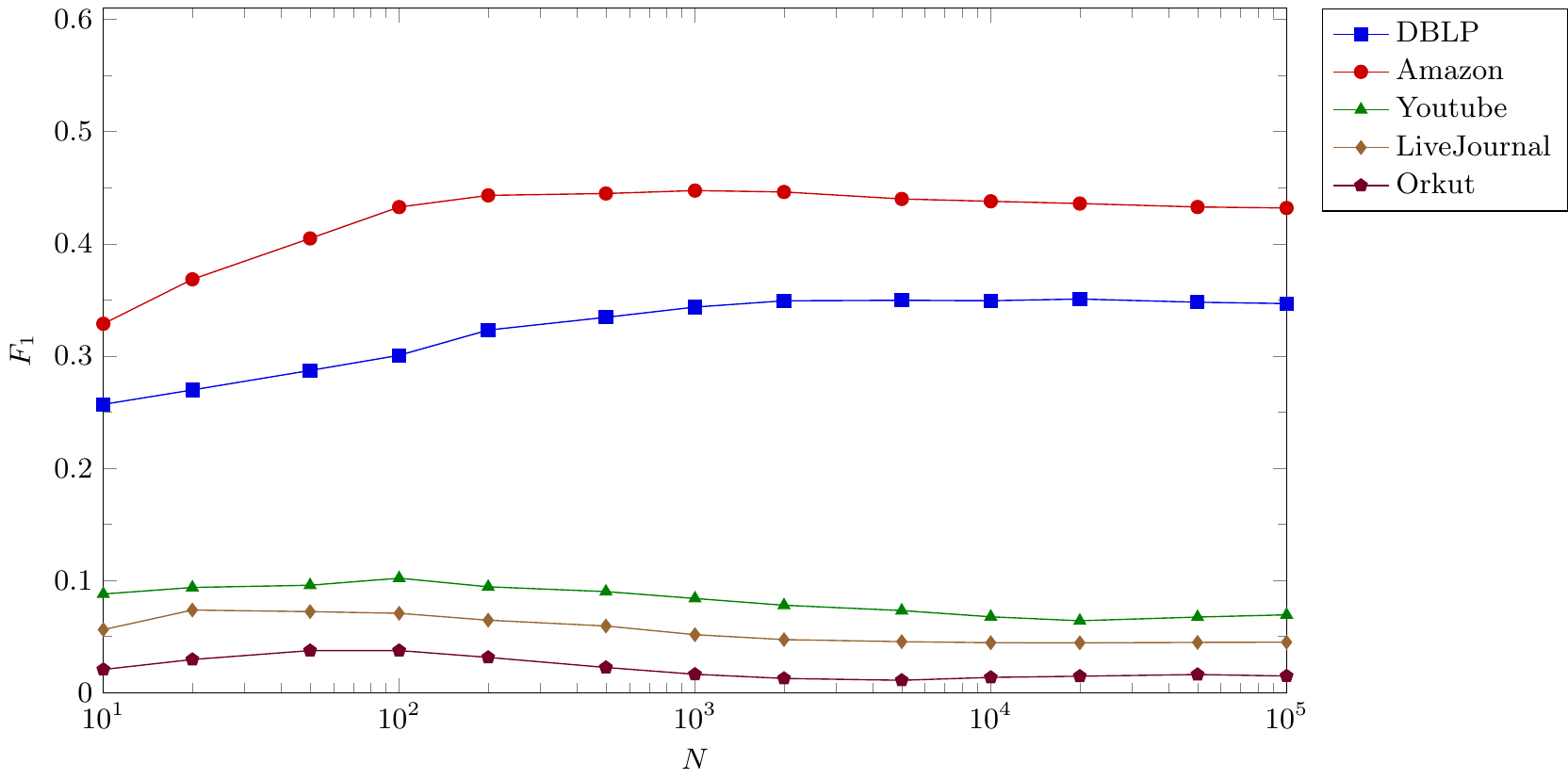}
  \caption{$F_1$ score as a function of the $N$ parameter with the aDCBM method.}
  \label{fig:N-f1}
\end{figure*}

\begin{figure*}
  \centering
  \includegraphics{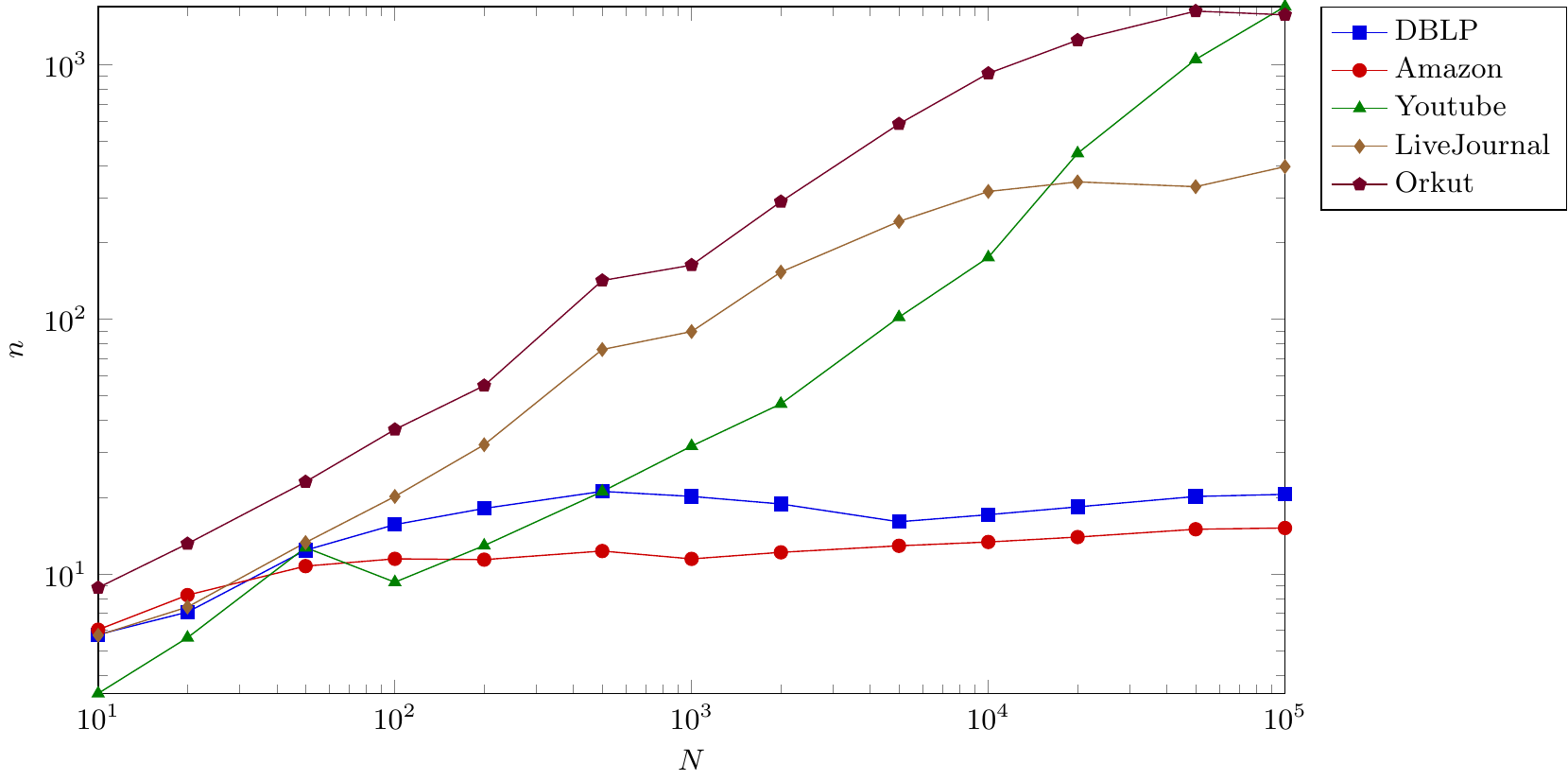}
  \caption{Mean community size as a function of the $N$ parameter with the aDCBM method.}
  \label{fig:N-size}
\end{figure*}

The tables and figures below include additional results: 
\begin{itemize}
  \item \tblref{tbl:moredatasets}: Statistics on the additional datasets.
  \item \tblref{tbl:size}: Mean size of the recovered community.
  \item \tblref{tbl:precision}: Mean precision of the recovered community.
  \item \tblref{tbl:recall}: Mean recall of the recovered community.
  \item \tblref{tbl:runtime}: Runtime of the community detection algorithms.
  \item \figref{fig:N-f1}: $F_1$ score on SNAP datasets as a function of the $N$ parameter.
  \item \figref{fig:N-size}: Mean size of the recovered community as a function of $N$ parameter.
\end{itemize}

\begin{table*}[ht]
  \caption{
    Overview of the additional datasets.
  }
  \label{tbl:moredatasets}
  \begin{center}
\begin{tabular}{lrr@{\hspace*{5mm}}r@{\hspace*{3mm}}r}
\hlinetop
Dataset & \#node & \#edge & \#comm & $\overline{|c|}$ \\
\hlinemid
LFR (mu=0.1) &    5000 &    25111 &    101 & 49.5\\
LFR (mu=0.2) &    5000 &    25124 &    101 & 49.5\\
LFR (mu=0.3) &    5000 &    25125 &    101 & 49.5\\
LFR (mu=0.4) &    5000 &    25127 &    101 & 49.5\\
LFR (mu=0.5) &    5000 &    25118 &    101 & 49.5\\
LFR (mu=0.6) &    5000 &    25127 &    101 & 49.5\\
\hlinemid
Amazon      &  334863 &   925872 & 151037 & 19.4\\
DBLP        &  317080 &  1049866 &  13477 & 53.4\\
Youtube     & 1134890 &  2987624 &   8385 & 13.5\\
LiveJournal & 3997962 & 34681189 & 287512 & 22.3\\
Orkut       & 3072441 & 117185083 & 6288363 & 14.2\\
\hlinemid
Amazon (top 5000) &  334863 &   925872 &   5000 & 13.5\\
DBLP (top 5000) &  317080 &  1049866 &   5000 & 22.4\\
Youtube (top 5000) & 1134890 &  2987624 &   5000 & 14.6\\
LiveJournal (top 5000) & 3997962 & 34681189 &   5000 & 27.8\\
Orkut (top 5000) & 3072441 & 117185083 &   5000 & 215.7\\
\hlinemid
Karate      &      34 &       78 &      2 & 17.0\\
Football    &     115 &      613 &     12 & 9.6\\
Pol.Blogs   &    1490 &    16715 &      2 & 745.0\\
Pol.Books   &     105 &      441 &      3 & 35.0\\
Flickr      &   35313 &  3017530 &    171 & 4336.1\\
\hlinebot
\end{tabular}
  \end{center}
\end{table*}

\begin{table*}[ht]
  \caption{Mean size of the recovered communities.
  }
  \medskip
  \small
\begin{tabular}{l|rr|rrr|rrrr}
\hlinetop
 & \multicolumn{2}{|c|}{Global methods} & \multicolumn{7}{|c}{Local methods} \\
\noalign{\tblskip}
Dataset & gSBM & gDCBM & aSBM & aDCBM & aDCBM-1k & {\YL} & LEMON & {\HK} & {\PPR} \\
\hlinemid
LFR (mu=0.1) & 48.9 & 49.0 & 20.5 & 45.7 & 44.2 & 9.4 & 50.2 & 159.0 & 1518.9\\
LFR (mu=0.2) & 48.7 & 49.0 & 19.7 & 42.1 & 39.7 & 8.4 & 49.2 & 2264.6 & 2155.1\\
LFR (mu=0.3) & 46.2 & 49.0 & 18.1 & 40.9 & 32.4 & 5.9 & 43.4 & 2410.0 & 2366.2\\
LFR (mu=0.4) & 42.9 & 48.9 & 16.1 & 30.8 & 21.9 & 5.2 & 42.9 & 2398.3 & 2323.2\\
LFR (mu=0.5) & 30.5 & 45.9 & 13.4 & 20.6 & 14.3 & 4.8 & 49.5 & 2412.0 & 2294.6\\
LFR (mu=0.6) & 16.3 & 20.7 & 10.5 & 13.6 & 9.9 & 4.9 & 56.8 & 2412.8 & 2269.3\\
\hlinemid
Amazon & 23.9 & 33.6 & 11.1 & 16.3 & 11.5 & 6.4 & 41.9 & 88.8 & 20819.9\\
DBLP & 27.6 & 39.8 & 11.2 & 21.3 & 20.2 & 6.0 & 46.2 & 55.0 & 24495.0\\
Youtube & 1283.5 & 38.0 & 31.1 & 78.6 & 31.8 & 9.3 & 53.2 & 147.9 & 20955.5\\
LiveJournal & 375.0 & 253.2 & 82.1 & 107.6 & 89.5 & 10.8 & - & 153.2 & 3428.7\\
Orkut & 2089.9 & 1288.8 & 367.5 & 283.8 & 163.0 & 11.1 & - & 212.0 & 1634.0\\
\hlinemid
Amazon (top 5000) & 15.4 & 24.5 & 7.6 & 10.4 & 9.6 & 7.1 & 40.0 & 28.3 & 4828.6\\
DBLP (top 5000) & 24.8 & 37.7 & 8.8 & 15.6 & 12.3 & 6.2 & 40.0 & 46.8 & 9506.8\\
Youtube (top 5000) & 434.0 & 42.0 & 30.6 & 42.6 & 26.3 & 6.9 & 55.3 & 113.9 & 10867.7\\
LiveJournal (top 5000) & 98.5 & 158.1 & 23.7 & 48.7 & 36.6 & 13.0 & - & 100.0 & 549.8\\
Orkut (top 5000) & 973.5 & 1272.3 & 181.5 & 219.9 & 202.6 & 9.1 & - & 264.8 & 1607.4\\
\hlinemid
Karate & 2.8 & 2.0 & 4.6 & 6.2 & 15.3 & 8.8 & 15.0 & 16.7 & 17.1\\
Football & 8.9 & 9.8 & 9.5 & 10.2 & 10.8 & 8.8 & 44.9 & 40.5 & 56.5\\
Pol.Blogs & 100.7 & 18.0 & 45.1 & 20.3 & 97.9 & 7.2 & 68.2 & 492.7 & 1051.1\\
Pol.Books & 10.7 & 21.1 & 12.3 & 9.1 & 25.2 & 6.7 & 43.3 & 49.3 & 53.4\\
Flickr & 2796.9 & 608.5 & 1054.0 & 222.5 & 436.9 & 12.9 & 75.4 & 174.2 & 1158.1\\
\hlinebot
\end{tabular}
  \label{tbl:size}
\end{table*}

\begin{table*}[ht]
  \caption{Mean precision of the recovered communities.
  }
  \medskip
  \small
\begin{tabular}{l|cc|ccc|cccc}
\hlinetop
 & \multicolumn{2}{|c|}{Global methods} & \multicolumn{7}{|c}{Local methods} \\
\noalign{\tblskip}
Dataset & gSBM & gDCBM & aSBM & aDCBM & aDCBM-1k & {\YL} & LEMON & {\HK} & {\PPR} \\
\hlinemid
LFR (mu=0.1) & \textbf{1.000} & \textbf{1.000} & 0.992 & 0.924 & 0.919 & 0.929 & 0.928 & 0.862 & 0.330\\
LFR (mu=0.2) & \textbf{1.000} & \textbf{1.000} & 0.982 & 0.827 & 0.840 & 0.823 & 0.870 & 0.069 & 0.114\\
LFR (mu=0.3) & 0.971 & \textbf{0.997} & 0.945 & 0.826 & 0.792 & 0.669 & 0.690 & 0.020 & 0.020\\
LFR (mu=0.4) & 0.951 & \textbf{0.988} & 0.898 & 0.725 & 0.651 & 0.569 & 0.607 & 0.020 & 0.021\\
LFR (mu=0.5) & \textbf{0.890} & \textbf{0.915} & 0.800 & 0.562 & 0.468 & 0.465 & 0.446 & 0.020 & 0.021\\
LFR (mu=0.6) & \textbf{0.647} & 0.577 & \textbf{0.644} & 0.236 & 0.225 & 0.387 & 0.266 & 0.019 & 0.020\\
\hlinemid
Amazon & 0.256 & 0.175 & 0.385 & 0.376 & \textbf{0.409} & \textbf{0.422} & 0.157 & 0.166 & 0.088\\
DBLP & 0.231 & 0.158 & \textbf{0.497} & 0.346 & 0.378 & 0.372 & 0.167 & 0.161 & 0.153\\
Youtube & 0.023 & 0.035 & \textbf{0.082} & 0.058 & 0.066 & \textbf{0.086} & 0.047 & 0.024 & 0.037\\
LiveJournal & 0.028 & 0.016 & \textbf{0.055} & 0.033 & 0.039 & \textbf{0.059} & - & 0.021 & 0.025\\
Orkut & 0.005 & 0.004 & 0.012 & 0.008 & 0.010 & \textbf{0.032} & - & 0.020 & 0.012\\
\hlinemid
Amazon (top 5000) & 0.743 & 0.546 & \textbf{0.977} & 0.949 & 0.959 & 0.928 & 0.317 & 0.649 & 0.469\\
DBLP (top 5000) & 0.343 & 0.233 & \textbf{0.740} & 0.548 & 0.594 & 0.544 & 0.229 & 0.274 & 0.260\\
Youtube (top 5000) & 0.052 & 0.052 & 0.122 & 0.149 & 0.139 & \textbf{0.226} & 0.062 & 0.045 & 0.080\\
LiveJournal (top 5000) & 0.405 & 0.275 & \textbf{0.696} & 0.641 & 0.623 & 0.592 & - & 0.479 & 0.400\\
Orkut (top 5000) & 0.173 & 0.136 & \textbf{0.377} & 0.299 & 0.140 & 0.235 & - & 0.278 & 0.207\\
\hlinemid
Karate & 0.607 & 0.427 & \textbf{0.958} & \textbf{0.961} & 0.872 & 0.913 & 0.732 & 0.813 & 0.909\\
Football & \textbf{0.825} & \textbf{0.818} & 0.755 & 0.686 & 0.768 & \textbf{0.828} & 0.204 & 0.371 & 0.154\\
Pol.Blogs & 0.603 & 0.751 & \textbf{0.806} & 0.732 & 0.670 & \textbf{0.776} & \textbf{0.795} & \textbf{0.765} & 0.433\\
Pol.Books & 0.625 & 0.610 & 0.641 & 0.624 & 0.616 & \textbf{0.699} & 0.596 & 0.587 & 0.586\\
Flickr & 0.271 & 0.242 & \textbf{0.395} & 0.098 & 0.271 & 0.125 & 0.294 & 0.201 & 0.222\\
\hlinebot
\end{tabular}
  \label{tbl:precision}
\end{table*}

\begin{table*}[ht]
  \caption{Mean recall of the recovered communities.
  }
  \medskip
  \small
\begin{tabular}{l|cc|ccc|cccc}
\hlinetop
 & \multicolumn{2}{|c|}{Global methods} & \multicolumn{7}{|c}{Local methods} \\
\noalign{\tblskip}
Dataset & gSBM & gDCBM & aSBM & aDCBM & aDCBM-1k & {\YL} & LEMON & {\HK} & {\PPR} \\
\hlinemid
LFR (mu=0.1) & 0.999 & \textbf{1.000} & 0.453 & 0.909 & 0.892 & 0.232 & 0.940 & \textbf{0.995} & \textbf{1.000}\\
LFR (mu=0.2) & 0.997 & \textbf{1.000} & 0.425 & 0.811 & 0.792 & 0.218 & 0.865 & \textbf{1.000} & \textbf{1.000}\\
LFR (mu=0.3) & 0.957 & \textbf{0.997} & 0.380 & 0.794 & 0.713 & 0.114 & 0.561 & \textbf{1.000} & \textbf{1.000}\\
LFR (mu=0.4) & 0.907 & 0.993 & 0.322 & 0.647 & 0.510 & 0.077 & 0.503 & \textbf{1.000} & \textbf{1.000}\\
LFR (mu=0.5) & 0.706 & 0.914 & 0.248 & 0.438 & 0.297 & 0.052 & 0.439 & \textbf{1.000} & \textbf{0.999}\\
LFR (mu=0.6) & 0.362 & 0.366 & 0.168 & 0.130 & 0.071 & 0.041 & 0.330 & \textbf{0.971} & \textbf{0.964}\\
\hlinemid
Amazon & 0.808 & 0.831 & 0.575 & 0.750 & 0.689 & 0.467 & 0.913 & 0.755 & \textbf{0.985}\\
DBLP & 0.543 & 0.593 & 0.421 & 0.495 & 0.416 & 0.223 & 0.701 & 0.483 & \textbf{0.853}\\
Youtube & 0.445 & 0.267 & 0.300 & 0.181 & 0.252 & 0.112 & 0.436 & 0.117 & \textbf{0.757}\\
LiveJournal & 0.346 & 0.345 & 0.286 & 0.172 & 0.260 & 0.081 & - & 0.193 & \textbf{0.661}\\
Orkut & 0.454 & 0.445 & 0.225 & 0.204 & 0.257 & 0.056 & - & 0.432 & \textbf{0.690}\\
\hlinemid
Amazon (top 5000) & 0.967 & 0.972 & 0.792 & 0.901 & 0.883 & 0.763 & 0.943 & 0.977 & \textbf{0.997}\\
DBLP (top 5000) & 0.811 & 0.856 & 0.649 & 0.737 & 0.667 & 0.396 & 0.897 & 0.723 & \textbf{0.939}\\
Youtube (top 5000) & 0.669 & 0.605 & 0.421 & 0.483 & 0.420 & 0.364 & 0.714 & 0.272 & \textbf{0.826}\\
LiveJournal (top 5000) & 0.932 & 0.938 & 0.707 & 0.892 & 0.803 & 0.554 & - & 0.910 & \textbf{0.962}\\
Orkut (top 5000) & 0.663 & \textbf{0.683} & 0.396 & 0.463 & 0.175 & 0.095 & - & 0.530 & \textbf{0.711}\\
\hlinemid
Karate & 0.101 & 0.060 & 0.241 & 0.317 & 0.721 & 0.446 & 0.641 & 0.794 & \textbf{0.913}\\
Football & 0.777 & 0.822 & 0.710 & 0.679 & 0.781 & 0.758 & 0.894 & \textbf{0.921} & \textbf{0.923}\\
Pol.Blogs & 0.129 & 0.020 & 0.055 & 0.023 & 0.069 & 0.008 & 0.084 & 0.593 & \textbf{0.707}\\
Pol.Books & 0.196 & 0.364 & 0.198 & 0.171 & 0.417 & 0.112 & 0.670 & 0.760 & \textbf{0.819}\\
Flickr & \textbf{0.259} & 0.089 & 0.127 & 0.049 & 0.046 & 0.012 & 0.022 & 0.042 & 0.125\\
\hlinebot
\end{tabular}
  \label{tbl:recall}
\end{table*}

\begin{table*}[ht]
  \caption{Mean runtime in seconds per community.
  Note that for the global methods a partition of the nodes is computed only once.
  }
  \medskip
  \small
\begin{tabular}{l|rr|rrr|rrrr}
\hlinetop
 & \multicolumn{2}{|c|}{Global methods} & \multicolumn{7}{|c}{Local methods} \\
\noalign{\tblskip}
Dataset & gSBM & gDCBM & aSBM & aDCBM & aDCBM-1k & {\YL} & LEMON & {\HK} & {\PPR} \\
\hlinemid
LFR (mu=0.1) & \textbf{0.017} & \textbf{0.017} & 0.024 & 0.024 & 0.030 & 0.128 & 0.113 & 0.047 & 1.012\\
LFR (mu=0.2) & \textbf{0.017} & \textbf{0.017} & 0.026 & 0.028 & 0.033 & 0.149 & 0.122 & 0.061 & 1.095\\
LFR (mu=0.3) & \textbf{0.018} & \textbf{0.018} & 0.029 & 0.032 & 0.035 & 0.069 & 0.047 & 0.034 & 0.649\\
LFR (mu=0.4) & \textbf{0.018} & \textbf{0.018} & 0.031 & 0.032 & 0.035 & 0.128 & 0.131 & 0.068 & 1.091\\
LFR (mu=0.5) & \textbf{0.017} & \textbf{0.018} & 0.032 & 0.033 & 0.035 & 0.139 & 0.132 & 0.069 & 1.123\\
LFR (mu=0.6) & \textbf{0.018} & \textbf{0.018} & 0.030 & 0.032 & 0.035 & 0.132 & 0.133 & 0.072 & 1.116\\
\hlinemid
Amazon & 0.036 & 0.035 & \textbf{0.022} & \textbf{0.023} & 0.029 & 4.418 & 1.564 & 0.031 & 2.803\\
DBLP & 0.032 & 0.034 & \textbf{0.026} & \textbf{0.027} & 0.035 & 5.799 & 1.757 & 0.039 & 1.405\\
Youtube & 0.059 & \textbf{0.055} & 0.854 & 1.138 & 0.200 & 18.975 & 4.959 & 0.074 & 0.514\\
LiveJournal & 0.130 & \textbf{0.116} & 0.726 & 0.684 & 0.508 & 189.899 & - & \textbf{0.115} & 0.312\\
Orkut & \textbf{0.095} & \textbf{0.095} & 7.551 & 4.923 & 3.487 & 1632.087 & - & 0.212 & 0.488\\
\hlinemid
Amazon (top 5000) & 0.034 & 0.032 & 0.020 & \textbf{0.018} & 0.027 & 2.629 & 1.559 & 0.022 & 1.643\\
DBLP (top 5000) & 0.034 & 0.036 & \textbf{0.022} & \textbf{0.023} & 0.028 & 5.637 & 1.736 & 0.036 & 1.319\\
Youtube (top 5000) & \textbf{0.056} & \textbf{0.056} & 0.707 & 0.398 & 0.146 & 18.295 & 4.855 & 0.078 & 0.520\\
LiveJournal (top 5000) & 0.124 & 0.118 & \textbf{0.111} & \textbf{0.095} & \textbf{0.095} & 101.696 & - & \textbf{0.101} & 0.446\\
Orkut (top 5000) & 0.101 & \textbf{0.088} & 1.960 & 1.774 & 2.144 & 1542.724 & - & 0.228 & 0.602\\
\hlinemid
Karate & 0.016 & 0.016 & 0.017 & 0.015 & 0.019 & 0.009 & 0.008 & \textbf{0.007} & 0.017\\
Football & 0.016 & 0.016 & 0.017 & 0.016 & 0.020 & 0.019 & 0.012 & \textbf{0.006} & 0.046\\
Pol.Blogs & 0.016 & 0.016 & 0.044 & 0.074 & 0.069 & 0.055 & 0.038 & \textbf{0.015} & 0.365\\
Pol.Books & 0.016 & 0.016 & 0.018 & 0.016 & 0.021 & 0.011 & 0.008 & \textbf{0.006} & 0.040\\
Flickr & 0.020 & \textbf{0.019} & 3.701 & 4.015 & 2.017 & 25.639 & 11.929 & 0.045 & 0.133\\
\hlinebot
\end{tabular}
  \label{tbl:runtime}
\end{table*}

\end{document}